\documentclass{iopart}
\input epsf

\def  \api	{\langle\pi\rangle}
\def  \kpi	{K/\pi}
\def  \kppi	{K^+/\pi}
\def  \kmpi	{K^-/\pi}
\def  \amt	{\langle m_{\perp} \rangle}
\def  \npart    {N_{\rm part}}
\def  \snn	{\sqrt{s_{_{\rm NN}}}}
\def  \pbar	{\overline{\rm p}}

\def  \qt	{Q_{\perp}}

\begin{document}
\title{Systematics of mid-rapidity $\kmpi$ ratio in heavy-ion collisions}
\author{Fuqiang Wang}
\address{Department of Physics, Purdue University, 
	West Lafayette, IN 47907, USA}
\address{E-mail: fqwang@physics.purdue.edu}

\begin{abstract}
It is observed that $\kmpi$ in A+A and possibly p+p and $\pbar$+p collisions 
follows interesting systematics in $\omega$, 
the pion transverse energy per unit of rapidity and transverse overlap area. 
The systematics show a linear increase of $\kmpi$ with $\omega$ in the AGS
and SPS energy regime and a saturation at RHIC energy.
The systematics indicate that $\omega$ might be the relevant variable 
underlying $\kmpi$.
At high energy, the $\omega$ variable is related to 
the gluon saturation scale in high density QCD, and perhaps to 
the initial energy density in the Bjorken picture.
\end{abstract}



Lattice QCD predicts that at sufficiently high energy density 
matter should be in a state of deconfined quarks and gluons~\cite{Karsch}. 
It has been suggested long ago~\cite{Rafelski} 
that strangeness production is a sensitive probe to such a deconfined state.
One of the common observables to search for
strangeness enhancement is the $K/\pi$ ratio.

\begin{figure}[hbt]
\hfill{
\epsfxsize=0.75\textwidth\epsfbox[50 120 520 420]{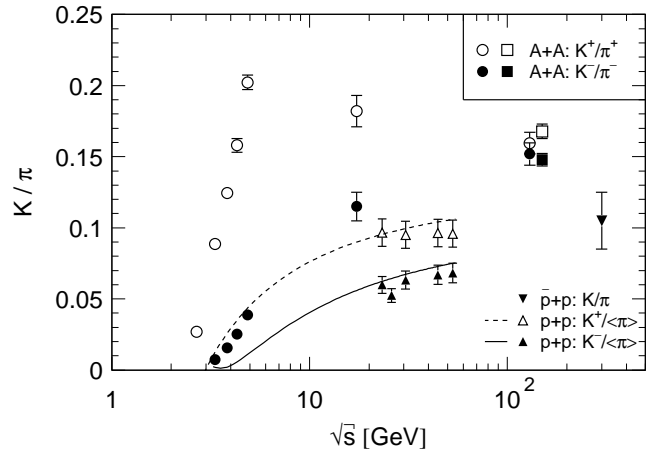}}
\caption{Mid-rapidity $K/\pi$ in central A+A~[3-5] and minimum bias p+p~[7]
and $\pbar$+p~[8] collisions as a function of the collision energy, $\snn$. 
Errors are statistical. The curves are parameterizations to low energy 
p+p data. The SPS/NA49 data ($\snn =17$~GeV) 
and RHIC/STAR data ($\snn = 130$~GeV) are preliminary. 
The two different measurements from STAR are displaced in $\snn$ for clarity.}
\label{kpi_roots}
\end{figure}

Strangeness production and $\kpi$ have been intensively studied in heavy-ion 
collisions at the AGS~\cite{AGS}, SPS~\cite{SPS}, and RHIC~\cite{RHIC}. 
Figure~\ref{kpi_roots} compiles $\kpi$ in central heavy-ion (A+A)
collisions~\cite{AGS,SPS,RHIC} as a function of 
the collision energy, $\snn$.
$\kmpi$ steadily increases with $\snn$, 
while $\kppi$ sharply increases at low energies.
The addition of the RHIC $\kppi$ measurement clearly demonstrates that 
$\kppi$ drops at high energies. 
A maximum $\kppi$ value is reached at about $\snn\approx 10$~GeV.
This behavior of $\kppi$ lies in the net-baryon density which changes 
significantly with $\snn$, as noted previously~\cite{netBaryonDensity}.
It is instructive to consider the two possible kaon production mechanisms: 
pair production of $K$ and $\overline{K}$ which is sensitive to $\snn$, 
and associated production of $K$ ($\overline{K}$) with a hyperon 
(antihyperon) which is sensitive to the net-baryon density.
In other words, a maximum in $\kppi$ results from a dropping 
net-baryon density with energy and an increasing production rate.
For comparison, Fig.~\ref{kpi_roots} also shows data 
from p+p~\cite{pp} and $\pbar$+p~\cite{pbarp}.
Enhancement in $\kmpi$ from elementary p+p to A+A collisions is about 
a factor of 2 and is similar at the SPS and RHIC, while that in $\kppi$ 
is larger at lower energies due to the large net-baryon density in A+A.
As $\kppi$ is complicated by the net-baryon density, 
the rest of the paper will focus on $\kmpi$.

\begin{figure}[hbt]
\hfill{\epsfxsize=0.7\textwidth\epsfbox[20 70 600 480]{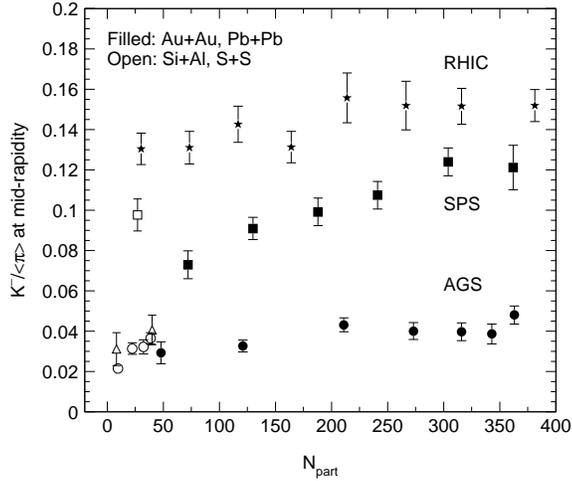}}
\caption{Mid-rapidity $\kmpi$ in heavy-ion collisions~[3-5] as a function of
the number of participants, $\npart$. Errors are statistical.
The data from RHIC/STAR (filled stars), SPS/NA49 (filled squares), 
AGS/E859 (open circles) and AGS/E866 (filled circles) are preliminary. 
The RHIC $\npart$ was obtained from a Glauber model calculation.}
\label{kmpi_npart}
\end{figure}

Experiments at the AGS and SPS have studied the centrality dependence 
of kaon production in A+A. 
Figure~\ref{kmpi_npart} shows $\kmpi$ as a function of the 
number of participants, $\npart$. 
%
$\kmpi$ increases with $\npart$ within the same collision system, 
however, differs in different systems at the same value of $\npart$, 
indicating that $\npart$ is not an appropriate variable to describe $\kmpi$.
This has been noted and emphasized before~\cite{AGS,SPS}. 

As demonstrated thus far, 
$\kmpi$ depends on both the collision energy and centrality. 
To identify the possible underlying physics is the goal of the
present study. To this end, we first note that
strangeness production may be enhanced due to the fast and energetically
favorable process of gluon-gluon fusion into strange quark-antiquark pairs, 
and therefore may be sensitive to the initial gluon density.
%
At high energies, the bulk of mid-rapidity hadrons are products of gluons 
of transverse momentum $\qt$$\sim$1~GeV/$c$ in a pseudo-rapidity range 
$\Delta\eta$$\sim$1~\cite{smallx} (i.e. $Q_z\sim\qt$). 
From Heisenberg uncertainty principle, these gluons 
occupy a longitudinal size of $h/Q_z \sim 1$~fm,
which is larger than the Lorentz contracted nucleus size at RHIC energy.
Each gluon therefore sees all the nucleons in the longitudinal direction
(they overlap longitudinally); the relevant quantity should be 
the gluon transverse area density~\cite{smallx}.
When the gluon density, $xG(x,\qt^2)\npart$, at Bjorken $x\sim 2Q_z/\snn$
is large such that gluons overlap in a volume of 
the gluon transverse size, $h/\qt$, 
the gluons will recombine populating the large $\qt$ region, 
resulting in a saturation of gluon density with $\qt<Q_s$. 
The saturation scale $Q_s$, based on the above uncertainty principle argument,
is $Q_s^2 = 4\pi^2 xG(x,Q_s^2) \npart / \pi r^2$,
where $\pi r^2$ is the transverse overlap area of the colliding nuclei.
That derived from QCD~\cite{saturation} is 
$Q_s^2 = 4\pi^2 N_c/(N_c-1) \alpha_s(Q_s^2) xG(x,Q_s^2) \rho_{\rm part}$, 
where the number of colors $N_c$=3, 
the strong coupling constant $\alpha_s$$\sim$0.6, 
and $\rho_{\rm part}$ is the participant density in the transverse plane.
At RHIC energies $Q_s$ is on the order of 1--2~GeV/$c$.

It has been argued that particle production at RHIC (and perhaps at SPS)
is dominated by the gluon saturation region~\cite{saturation}.
While number of constituents is not conserved during the hadronization
process, the total transverse energy of gluons ($\propto Q_s^3$) 
may be better preserved in the final hadrons.
Motivated by the gluon saturation picture, 
we introduce a new, experimental variable:
\begin{equation}
\omega = \amt_{\rm cent} \cdot 
	\frac{3\,(dN_{\api}/dy)_{\rm cent}}{\pi R^2} \cdot
	\left[\frac{dN_{\api}/dy}{(dN_{\api}/dy)_{\rm cent}}\right]^{1/3} \; .
\label{eq:w}
\end{equation}
Since pion multiplicity is approximately proportional to the overlap volume 
(or $\npart$), and since $\amt$ is approximately constant over the measured
centralities, $\omega$ is an approximate measure of the pion transverse
energy per unit of rapidity and transverse overlap area: 
\begin{equation}
\omega \approx	\amt \cdot \frac{3\,dN_{\api}/dy}{\pi r^2} \propto Q_s^3 \; .
\label{eq:w_approx}
\end{equation}
Here $\amt$ and $dN_{\api}/dy$ are the mid-rapidity pion mean transverse mass 
and multiplicity density, respectively, 
and the subscript `cent' stands for the most central collisions; 
$r$ is the transverse radial size of nuclear overlap, and 
$R$=$1.12(\npart/2)^{1/3}_{\rm cent}$ is that for the most central collisions.
The first factor on the r.h.s. of Eq.~(\ref{eq:w}) 
quantifies transverse energy production; 
The second factor depends on the collision energy: the larger the energy,
the smaller the $x$ probed by mid-rapidity hadrons, 
and the larger the gluon density; 
The third factor defines centrality: 
the more nucleons in the longitudinal direction, 
the larger the gluon density in the transverse plane.
The second and third factors combined, as in Eq.~(\ref{eq:w_approx}), 
scales with the transverse area gluon density.
The $\omega$ variable therefore unifies all three effects and 
is of use in studying systematics in heavy ion collisions.

\begin{figure}[hbt]
\hfill{\epsfxsize=0.8\textwidth\epsfbox[0 60 560 480]{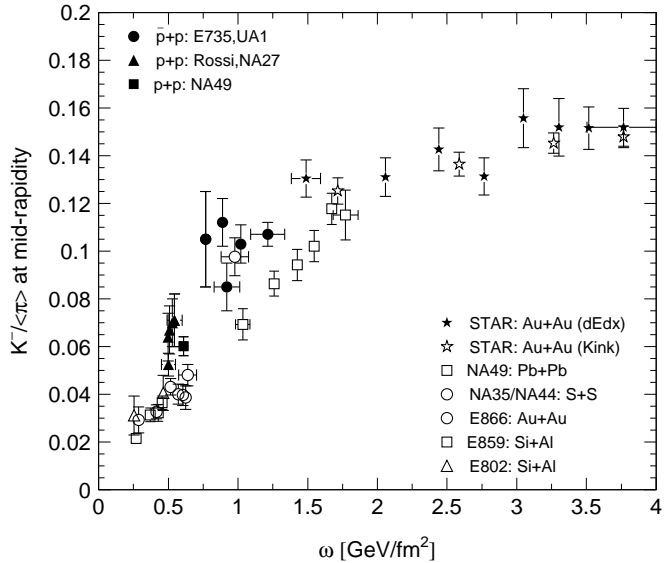}}
\caption{Mid-rapidity $\kmpi$ ratio in heavy-ion collisions~[3-5] 
as well as p+p~[7] and $\pbar$+p~[8] collisions as a function of $\omega$. 
Systematic errors on $\omega$ are shown for selected points.
Errors on $\kmpi$ are statistical.
The STAR, NA49, E859 and E866 data are preliminary.}
\label{kmpi_w}
\end{figure}

Figure~\ref{kmpi_w} shows the mid-rapidity $\kmpi$ as a function of $\omega$ 
in heavy-ion collisions~\cite{AGS,SPS,RHIC} including 
different collision systems, energies, and centralities.
The data seem to follow a trend: 
a linear increase of $\kmpi$ with $\omega$ in the AGS and SPS energy regime 
and a saturation at RHIC with further increase in $\omega$.
The observed trend could imply that $\omega$ may be the relevant variable 
underlying $\kmpi$. 
An interesting study in terms of the participant density $\rho_{\rm part}$ 
can be found in~\cite{Polish}.

It is worthwhile to emphasize that the $\omega$ defined in Eq.~(\ref{eq:w}) 
is an experimental variable. 
The interpretation, however, needs a leap of faith. 
In the gluon saturation picture, it is possible that 
the initial gluon density is saturated at the SPS and RHIC energies.
As the saturation scale $Q_s \propto \omega^{1/3}$ becomes large,
the difference between kaon and pion masses becomes less important, 
resulting in a roughly constant $\kmpi$.
Gluon saturation should be irrelevant at AGS energies, 
as gluons can be distinguished longitudinally and 
quark contribution to particle production is significant. 
However, the fact that Si+Al and Au+Au data are on top of each other
in Fig.~\ref{kmpi_w} indicates that $\omega$ may be the relevant quantity
for $\kmpi$ at the AGS, although the interpretation may be different from 
that at high energies.

A quick examination of Eq.~(\ref{eq:w_approx}) reveals that $\omega$ is 
approximately equal to the Bjorken estimate~\cite{Bjorken} 
of the initial energy density ($\epsilon$) times the formation time ($\tau$). 
Taking $\tau$ as constant, Fig.~\ref{kmpi_w} implies that $\kmpi$,
while is increasing with $\epsilon$ at the SPS, becomes insensitive to
$\epsilon$ when $\epsilon$ becomes large at RHIC. 
A possible explanation is that the chemical freeze-out condition 
(which fixes $\kmpi$) is sensitive to $\epsilon$ at SPS, 
while at RHIC $\epsilon$ is no longer relevant to 
the (later) chemical freeze-out condition.
We note, as $\tau$ likely decreases with the collision energy and 
some of the transverse energy is carried by particles other than pions,
the above line of argument is at best qualitative.

Strangeness enhancement is often referred to as an enhancement in 
$\kpi$ in A+A with respect to minimum bias p+p at the same energy 
(e.g. as in Fig.~\ref{kpi_roots}). 
However, this may not be justified as energy is not the only relevant 
variable. If $\omega$ is indeed the relevant variable, then A+A should have
a larger $\omega$ value than the same energy p+p due to the $A^{1/3}$ factor.
To calculate $\omega$ for p+p and $\pbar$+p by Eq.~(\ref{eq:w_approx}),
we use $r=0.8$~fm, the proton size from the MIT bag model~\cite{Wong}. 
Figure~\ref{kmpi_w} shows in filled points $\kmpi$ 
in minimum bias p+p and $\pbar$+p as a function of $\omega$.
The p+p and $\pbar$+p data are, surprisingly, not far away from the
systematics observed in A+A. 
We note that the $r$ value used is rather model dependent;
a choice of $r$=0.6--0.7~fm would bring the p+p and $\pbar$+p data points 
on top of A+A.


In summary, an interesting observation is made for the mid-rapidity 
$\kmpi$ ratio in heavy-ion collisions as a function of the experimental 
$\omega$ variable, the pion transverse energy per unit of rapidity and 
transverse overlap area, as defined in Eq.~(\ref{eq:w}). 
At high collision energies, the variable may be related to 
the gluon saturation scale in high density QCD,
and/or the Bjorken estimate of initial energy density times formation time.
It is observed that $\kmpi$ increases linearly with $\omega$ 
in heavy-ion collisions at low energies and saturates at RHIC.
Whether or not the elementary p+p and $\pbar$+p data follow the 
same systematics depends on the choice of the proton size 
which is less certain than the transverse size of a heavy-ion collision.
To further establish the observed systematics, low energy RHIC data, 
overlapping with the existing SPS and RHIC data in $\omega$, are essential.

\section*{Acknowledgments}

I thank Drs. J. Dunlop, A. Hirsch, R. Scharenberg, F. Sikler, 
N. Xu, and Z. Xu for fruitful discussions.
This work was supported by the U.S. Department of Energy under 
contract DE-FG02-88ER40412.

\end{document}